# Maximum Likelihood Estimation of Triangular and Polygonal Distributions

Hien D. Nguyen and Geoffrey J. McLachlan

February 14, 2016

School of Mathematics and Physics, University of Queensland

## Abstract

Triangular distributions are a well-known class of distributions that are often used as elementary example of a probability model. In the past, enumeration and order statistic-based methods have been suggested for the maximum likelihood (ML) estimation of such distributions. A novel parametrization of triangular distributions is presented. The parametrization allows for the construction of an MM (minorization–maximization) algorithm for the ML estimation of triangular distributions. The algorithm is shown to both monotonically increase the likelihood evaluations, and be globally convergent. Using the parametrization is then applied to construct an MM algorithm for the ML estimation of polygonal distributions. This algorithm is shown to have the same numerical properties as that of the triangular distribution. Numerical simulation are provided to demonstrate the performances of the new algorithms against established enumeration and order statistics-based methods.

## 1 Introduction

Triangular distributions are a class of continuous distributions that are often used as an elementary example in first-level probability and statistics courses due to their simple geometric forms; see Johnson et al. (1995, Sec. 26.9) and Forbes et al. (2011, Ch. 44) for characterizations of such distributions. Outside of the classroom, triangular distributions have been used to model task completion times for PERT (program evaluation and review technique) models, prices associated with orders placed by investors on single securities that are traded on the New York Stock Exchange, and hauling times in civil engineering data (see Kotz and van Dorp (2004, Ch. 1) and references within for details).

Let $X \in [0,1]$ be a random variable with probability density

$$f(x;\theta) = \begin{cases} 2x/\theta & \text{if } 0 \leq x \leq \theta, \\ 2(1-x)/(1-\theta) & \text{if } \theta < x \leq 1, \end{cases} \qquad (1)$$

where $\theta \in (0,1)$. Densities of form (1) are known as triangular distributions. Suppose that we observe realizations $x_1, ..., x_n$ of a random sample $X_1, ..., X_n$ from a distribution with density (1) with unknown population parameter $\theta_0$. We can define the likelihood and log-likelihood of such as sample as $\mathcal{L}_n(\theta) = \prod_{j=1}^{n} f(x_j; \theta)$ and $\ell_n(\theta) = \sum_{j=1}^{n} \log f(x_j; \theta)$ and estimate the parameter $\theta_0$ via the maximum likelihood (ML) estimator

$$\hat{\theta}_n = \arg\max_{\theta \in (0,1)} \mathcal{L}_n(\theta) = \arg\max_{\theta \in (0,1)} \ell_n(\theta). \qquad (2)$$

Unfortunately in the case of (1), the computation of (2) cannot be performed by solving the usual score equation, $\mathrm{d}\ell_n/\mathrm{d}\theta = 0$, for $\theta$, since (1) is not differentiable in $\theta$. In attempting to find an elementary ML estimator for (1), Oliver (1972) established that $\hat{\theta}_n = x_i$ for some $i = 1, ..., n$; see also Johnson and Kotz (1999). Furthermore, Oliver (1972) deduced that if the sample is ordered, such that $x_{(1)} < x_{(2)} < ... < x_{(n)}$, then $\hat{\theta}_n \in \Theta_n$, where

$$\Theta_n = \left\{ x_{(i)} : \frac{i-1}{n} < x_{(i)} < \frac{i}{n}, i = 1, ..., n \right\}.$$

As such, it is enough to find all of the sample observations that are in the set $\Theta_n$ and compute

$$\hat{\theta}_n = \arg\max_{\theta \in \Theta_n} \mathcal{L}_n(\theta). \qquad (3)$$

An analysis on the distribution of the number of observations in $\Theta_n$ for various $n$ and $\theta_0$ is conducted in Huang and Shen (2007). Here, it appears that asymptotically, on average, the number of observations in $\Theta_n$ is approximately two. Thus, very few evaluations of the likelihood function are needed when utilizing the method of Oliver (1972).

Although the argument for the use of (3) is compelling, we wish to establish an estimation algorithm for computing (2) that utilizes a root finding argument, as opposed to one that is based on ordering and sieving. In order to do so, we consider an alternative parametrization of (1) and construct an MM (minorization–maximization) algorithm for the maximization of its likelihood function; see Lange (2013, Ch. 8) regarding MM algorithms.

The MM paradigm has been used to construct iterative algorithms for the computation of difficult ML estimates in the past. For example, by Wu and Lange (2010) for the multivariate $t$ distribution and power series distributions, by Zhou and Zhang (2012) for the Dirichlet-multinomial distribution, and by Zhou and Lange (2010) for various other multivariate discrete distributions.

Furthermore, MM algorithms have been constructed in Hunter (2004) for the estimation of generalized Bradley-Terry models, and in Hunter and Li (2005) to perform variable selection in linear regression models. More recently, MM algorithms have been constructed in Nguyen and McLachlan (2015) to obtain ML estimates for mixtures of Gaussian distributions without using matrix operations, and in Nguyen and McLachlan (2016) to obtain ML estimates for Laplace mixtures of linear experts.

Apart from our construction of an MM algorithm for the ML estimation of the triangular distribution, we also extend our algorithm to the ML estimation for the polygonal distribution of Karlis and Xekalaki (2008). Additionally, we obtain global convergence results for our MM algorithms in both the triangular and polygonal distribution cases. Reports on some simulations performed to demonstrate the relative performances of our algorithms against established methods of estimation are also included. The article proceeds as follows.

In Section 2, a novel parametrization of the triangular distribution is introduced. In Section 3, the MM algorithm paradigm is introduced and an MM algorithm is constructed for the ML estimation of the triangular distribution. In Section 3, the polygonal distribution is introduced, along with an MM algorithm for its ML estimation. Reports from some simulation studies are presented in Section 4, and conclusions are drawn in Section 5. Supplementary simulation results are presented in the Appendix.

## 2 Alternative Parametrization of the Triangular Distribution

We seek to write (1) in the form

$$f(x) = \min\{g_1(x), g_2(x)\}, \tag{4}$$

for $x \in [0, 1]$, $g_1(x) = a_1 + b_1 x$ and $g_2(x) = a_2 + b_2 x$, for some $a_1, a_2, b_1, b_2 \in \mathbb{R}$.

Without loss of generality, suppose that $g_1$ is the left ray and $g_2$ is the right ray of the triangle; see Figure 1. This implies that $a_1 = 0$ since $g_1$ has a root at $x = 0$ and $b_1 > 0$, similarly, since $g_2$ must have a root at when $x = 1$, we have $a_2 = -b_2$ and $b_2 < 0$. Thus, $g_1(x) = b_1 x$ and $g_2(x) = -b_2 + b_2 x$.

Next, we find that $x^* = -b_2/(b_1 - b_2)$ solves the intercept equation $g_1(x^*) = g_2(x^*)$, which yields a modal density value of $f(x^*) = b_1 b_2 / (b_2 - b_1)$. Since $f$ is a probability density function, we require that the area under $f$ for $x \in [0, 1]$ must equal to one. Since $f$ forms a triangle with the $x$-axis over the domain, we have $b_1 b_2 / 2 (b_2 - b_1) = 1$ or $b_2 = -2b_1/(b_1 - 2)$, by the formula for the area of a triangle. Finally, we make the substitution $\beta = b_1$ and note $b_2 < 0$ to obtain the alternative parametrization of the triangular distribution

$$f(x;\beta) = \min\{\beta x, 2\beta(1-x)/(\beta-2)\}, \tag{5}$$

for $x \in [0, 1]$ and $\beta > 2$.

Note that the mode $x^*$ is equal to $\theta$ from (1), which we can use to map between the two parametrizations via the equation $\theta = 2/\beta$. Therefore, we have the symmetric triangular distribution when $\beta = 4$. Furthermore, using Karlis and Xekalaki (2008, Eq. 2.3), we have the moment equations

$$\mathbb{E}(X^r) = \frac{2\left(1 - 2^{k+1}/\beta^{k+1}\right)}{(r+1)(r+2)(1-2/\beta)}$$

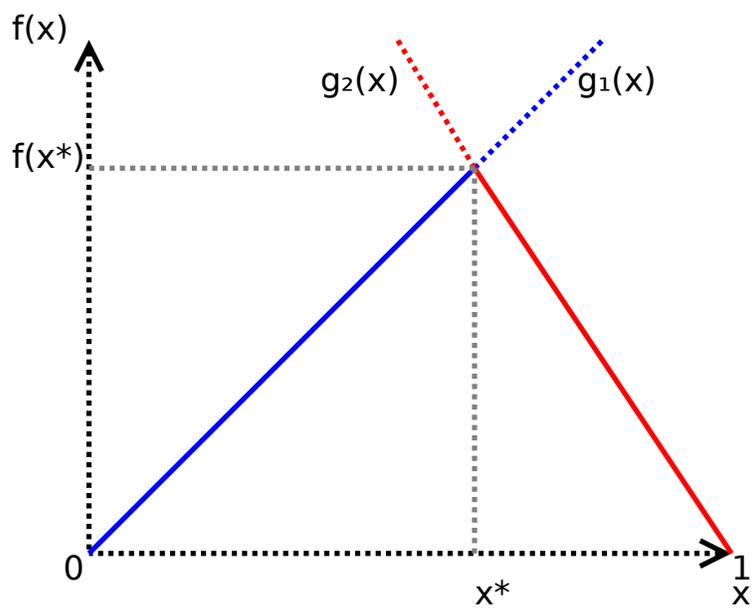

Figure 1: Reference diagram for the derivation of the alternative parametrization for the triangular distribution.

in terms of $\beta$, for each $k \in \mathbb{N}$, which yields the mean and variance of $X$: $\mathbb{E}(X) = 1/3 + 2/3\beta$ and $\text{var}(X) = 1/18 - 1/9\beta + 2/9\beta^2$, respectively.

## 3 Maximum Likelihood Estimation

Let $x_1, ..., x_n$ be a realization of a random sample $X_1, ..., X_n$ from a distribution with density (5) with unknown population parameter $\beta_0$. We can write the likelihood and log-likelihood for the sample as $\mathcal{L}_n(\beta) = \prod_{j=1}^n f(x_j; \beta)$ and

$$
\begin{aligned}
\ell_n(\beta) &= \sum_{j=1}^n \log \min \{\beta x_j, 2\beta(1-x_j)/(\beta-2)\} \\
&= \sum_{j=1}^n \log \left( \beta x_j + 2\frac{\beta(1-x_j)}{\beta-2} - \left| \frac{\beta(x_j\beta-2)}{\beta-2} \right| \right) - n \log 2, \quad (6)
\end{aligned}
$$

by noting that $\min\{x, y\} = (x + y - |x - y|)/2$ for any $x, y \in \mathbb{R}$. Due to the absolute value in (6), we cannot obtain the score equation, $\mathrm{d}\ell_n/\mathrm{d}\beta = 0$, in order to compute the ML estimate

$$
\hat{\beta}_n = \arg\max_{\beta>2} \mathcal{L}_n(\beta) = \arg\max_{\beta>2} \ell_n(\beta). \quad (7)
$$

As such, we seek to construct an MM algorithm for the iterative computation of (7).

### 3.1 MM Algorithms

Suppose that we wish to maximize some objective function $\mathcal{L}(\boldsymbol{\theta})$, where $\boldsymbol{\theta} \in \Theta \subset \mathbb{R}^d$. If we cannot obtain the maximizer of $\mathcal{L}$ directly (e.g. due to its lack of differentiability or difficult first-order conditions), then we can seek to minorize $\mathcal{L}$ over $\Theta$, instead. We say that $\mathcal{U}(\boldsymbol{\theta}; \boldsymbol{\psi})$ is a minorizer of $\mathcal{L}$ if $\mathcal{L}(\boldsymbol{\psi}) = \mathcal{U}(\boldsymbol{\psi}; \boldsymbol{\psi})$ and $\mathcal{L}(\boldsymbol{\theta}) \geq \mathcal{U}(\boldsymbol{\theta}; \boldsymbol{\psi})$, whenever $\boldsymbol{\theta} \neq \boldsymbol{\psi}$ and $\boldsymbol{\psi} \in \Theta$. Here, we say that $\mathcal{L}$ is minorized by $\mathcal{U}$.

Let $\boldsymbol{\theta}^{(0)}$ be some initial value of the MM algorithm, and let $\boldsymbol{\theta}^{(r)}$ denote the $r$th iterate of the algorithm. The MM algorithm can thus be defined by the update scheme

$$
\boldsymbol{\theta}^{(r+1)} = \arg\max_{\boldsymbol{\theta} \in \Theta} \mathcal{U}\left(\boldsymbol{\theta}; \boldsymbol{\theta}^{(r)}\right). \quad (8)
$$

By the property of the minorizer, and by definition (8), we have the inequalities

$$
\mathcal{L}\left(\boldsymbol{\theta}^{(r+1)}\right) \geq \mathcal{U}\left(\boldsymbol{\theta}^{(r+1)}; \boldsymbol{\theta}^{(r)}\right) \geq \mathcal{U}\left(\boldsymbol{\theta}^{(r)}; \boldsymbol{\theta}^{(r)}\right) = \mathcal{L}\left(\boldsymbol{\theta}^{(r)}\right).
$$

This implies that the sequence of objective function evaluations, $\mathcal{L}\left(\boldsymbol{\theta}^{(r)}\right)$, is monotonically increasing at each iteration. We now present a set of results that are useful in our MM algorithm construction.

**Fact 1.** *If $\Theta = \mathbb{R} \setminus \{0\}$, then $\mathcal{L}(\theta) = -|\theta|$ can be minorized by*

$$\mathcal{U}(\theta; \psi) = \frac{1}{2|\psi|}\left(\theta^2 + \psi^2\right).$$

**Fact 2.** *If $g$ is convex over $\Theta$, then $\mathcal{L}(\theta)$ can be minorized by*

$$\mathcal{U}(\theta; \psi) = g(\psi) + g'(\psi)(\theta - \psi).$$

**Fact 3.** *Let $\mathcal{F}$ be a strictly increasing function over $\Theta$. If $\mathcal{L}(\theta)$ is minorized by $\mathcal{U}(\theta; \psi)$, then $\mathcal{F}(\mathcal{L}(\theta))$ is minorized by $\mathcal{F}(\mathcal{U}(\theta; \psi))$.*

## 3.2 Algorithm Construction

In order to construct our MM algorithm, we first seek a minorizer for $2f(x;\beta)$, conditioned on the $r$th iterate $\beta^{(r)}$. An application of Fact 1 yields

$$\begin{aligned}
u_1^{(r)}(x;\beta) &= \beta x + 2\frac{\beta(1-x)}{\beta-2} - \frac{1}{2w^{(r)}(x)}\left(\left[\frac{\beta(x\beta-2)}{\beta-2}\right]^2 + \left[w^{(r)}(x)\right]^2\right) \\
&= \beta x + 2\frac{\beta(1-x)}{\beta-2} - \frac{1}{2w^{(r)}(x)}\frac{\beta^2(x\beta-2)^2}{(\beta-2)^2} - \frac{w^{(r)}(x)}{2},
\end{aligned}$$

where $w^{(r)}(x) = \left|\beta^{(r)}\left(x\beta^{(r)} - 2\right) / \left(\beta^{(r)} - 2\right)\right|$, by making the substitution $\theta = |\beta(x\beta - 2)/(\beta - 2)|$.

Using Fact 2, we minorize $u_1^{(r)}$ by

$$\begin{aligned}
u_2^{(r)}(x;\beta) &= \beta x + a^{(r)}(x) - b^{(r)}(x)\left(\beta - \beta^{(r)}\right) \\
&\quad - \frac{1}{2w^{(r)}(x)}\frac{\beta^2(x\beta-2)^2}{(\beta-2)^2} - \frac{w^{(r)}(x)}{2},
\end{aligned} \quad (9)$$

where $a^{(r)}(x) = 2\beta^{(r)}(1-x)/\left(\beta^{(r)} - 2\right)$ and $b^{(r)}(x) = 4(x-1)/\left(\beta^{(r)} - 2\right)^2$, by making the substitution $g(\beta) = 2\beta(1-x)/(\beta-2)$, and by noting that it is convex for $\beta > 2$ when $x \in [0,1]$.

We further note that (9) is concave in $\beta$ as it is composed of an affine combination of concave terms. The only nonlinear term in the expression is $\beta^2(x\beta - 2)^2/(\beta - 2)^2$, which can be shown to be convex as it is the product of three positive convex functions; thus its negative is concave. See Boyd and Vandenberghe (2004, Ch. 3) regarding the algebra of convex functions. An application of Fact 3, by making the substitutions of log for $\mathcal{F}$ and $\mathcal{L}(\beta) = u_2^{(r)}(x;\beta)$, yields the following result.

**Proposition 1.** *Conditioned on the $r$th iterate of the MM algorithm, the log-likelihood function (6) can be minorized by*

$$\mathcal{U}\left(\beta; \beta^{(r)}\right) = \sum_{j=1}^{n} \log u_2^{(r)}(x_j; \beta) - 2\log n, \quad (10)$$

and $\mathcal{U}\left(\beta;\beta^{(r)}\right)$ is a concave function in $\beta$.

Since (10) is concave, if the first-order condition $\mathrm{d}\mathcal{U}/\mathrm{d}\beta=0$ has a solution $\beta^{*}$, then $\beta^{*}$ must be the global maximum of (10). Unfortunately, we cannot obtain $\beta^{*}$ in closed form. Fortunately, we can obtain the derivative of (10), by noting that if $u\left(\theta\right)>0$ and differentiable, then $\mathrm{d}\log u/\mathrm{d}\theta=\left(\mathrm{d}u/\mathrm{d}\theta\right)/u\left(\theta\right)$, and by computing the derivative

$$\frac{\mathrm{d}u_2^{(r)}}{\mathrm{d}\beta}=x-b^{(r)}\left(x\right)-\frac{\beta\left(\beta x-2\right)^2}{w^{(r)}\left(x\right)\left(\beta-2\right)^2}-\frac{\beta^2 x\left(\beta x-2\right)}{w^{(r)}\left(x\right)\left(\beta-2\right)^2}+\frac{\beta^2\left(\beta x-2\right)^2}{w^{(r)}\left(x\right)\left(\beta-2\right)^3}.$$

A root finding algorithm such as the bisection method, or the method of Brent (1971) can then be used to compute $\beta^{*}$. Furthermore, a Newton algorithm can be constructed by noting that, $\mathrm{d}^2\log u/\mathrm{d}\theta^2=\left(\mathrm{d}^2\theta/\mathrm{d}\theta^2\right)/u\left(\theta\right)-\left(\mathrm{d}u/\mathrm{d}\theta\right)^2/u^2\left(\theta\right)$, and by computing the second derivative

$$\begin{aligned}\frac{\mathrm{d}^2 u_2^{(r)}}{\mathrm{d}\beta^2}&=-\frac{\left(\beta x-2\right)^2}{w^{(r)}\left(x\right)\left(\beta-2\right)^2}-4\frac{\beta x\left(\beta x-2\right)}{w^{(r)}\left(x\right)\left(\beta-2\right)^2}+4\frac{\beta\left(\beta x-2\right)^2}{w^{(r)}\left(x\right)\left(\beta-2\right)^3}\\&\quad-\frac{\beta^2 x^2}{w^{(r)}\left(x\right)\left(\beta-2\right)^2}+4\frac{\beta^2 x\left(\beta x-2\right)}{w^{(r)}\left(x\right)\left(\beta-2\right)^3}-3\frac{\beta^2\left(\beta x-2\right)^2}{w^{(r)}\left(x\right)\left(\beta-2\right)^4}.\end{aligned}$$

We can summarize the MM algorithm as follows. First, initialize the algorithm by some $\beta^{(0)}$. At the $(r+1)$th iteration of the algorithm, set $\beta^{(r+1)}=\beta^{*}$ where $\beta^{*}$ globally maximizes $\mathcal{U}\left(\beta;\beta^{(r)}\right)$.

### 3.3 Convergence Analysis

In general, the MM algorithm is iterated until convergence is achieved. In this article, we choose to use the convergence criterion $\ell\left(\beta^{(r+1)}\right)-\ell\left(\beta^{(r)}\right)<\epsilon$, where $\epsilon>0$ is a small constant; see Lange (2013, Sec. 11.5) regarding the relative merits of convergence criteria. Upon convergence, we declare the final iterate of the algorithm to be the ML estimator, and we denote it as $\hat{\beta}_n$.

Let $\beta^{(\infty)}=\lim_{r\to\infty}\beta^{(r)}$ be a finite limit point of the MM algorithm, starting from some initialization $\beta^{(0)}$ (or alternative, $\hat{\beta}_n\to\beta^{(\infty)}$ as $\epsilon\to 0$). We note that the MM algorithm fulfills the assumptions required to be defined as a successive lower-bound maximization (SLM) algorithm in the sense of Razaviyayn et al. (2013). As such, we can apply Razaviyayn et al. (2013, Thm. 1) directly to obtain the following result.

**Theorem 1.** *If $\beta^{(\infty)}$ is a limit point of the MM algorithm, for some initialization $\beta^{(0)}$, then $\beta^{(\infty)}$ is a local maximizer or inflection point of (6).*

Thus, the MM algorithm monotonically increases the likelihood evaluations and is globally convergent. Unfortunately, it is notable that (6) is highly multimodal and thus there is a great likelihood of the algorithm converging to a maximizer that does not globally maximize (6). Fortunately, initialization of the algorithm is simple since the method of moments estimator $\tilde{\beta}_n=2/\left(3\bar{x}_n-1\right)$ is an unbiased estimator of $\beta_0$, where $\bar{x}_n$ is the sample mean.

## 3.4 Time Complexity

We now consider a brief analysis of the time complexity of the MM algorithm, in terms of the sample size $n$, in comparison to alternative methods for computing (7). First, note that each Newton iteration involved in the MM algorithm requires a constant order of operation for each observation and thus each Newton iterate is of order $O(n)$. If we let $A_N^{[n]}$ and $A_{MM}^{[n]}$ be the average number Newton iterations taken to achieve a convergence level of $\epsilon$ and the average number of MM iterations taken to a achieve a convergence level of $\epsilon$, then the MM algorithm has a total complexity of order $O\left(nA_N^{[n]}A_{MM}^{[n]}\right)$.

Next, in order to compute (3), a sort operation is needed, which has a complexity order of $O(n \log n)$; notable sorting algorithms with this complexity are *heap sort* and *quick sort* (cf. Cormen et al. (2002, Pt. 2)). Furthermore, $O(n)$ evaluations are needed to obtain the set $\Theta_n$ and then $A_\Theta^{[n]}$ evaluation valuations of the likelihood function are required at the complexity of $O(n)$, where $A_\Theta^{[n]}$ is the average number of elements in $\Theta_n$. Thus, the complexity of computing (3) is of order $O\left(n\left(\log n + A_\Theta^{[n]}\right)\right)$. Lastly, it is not difficult to see that the naive computation the likelihood function at $\hat{\theta}_n = x_i$ for each $i$ has a time complexity order of $O(n^2)$.

If the observations of Huang and Shen (2007) are correct then computing (3) reduces to order $O(n \log n)$. Thus, the MM algorithm would be competitive to (3) if $A_N^{[n]}A_{MM}^{[n]}$ grows slower than $\log n$.

## 4 Polygonal Distributions

Let $X \in [0, 1]$ be a random variable with probability density function

$$f(x; \boldsymbol{\psi}) = \sum_{i=1}^{g} \pi_i f(x; \beta_i), \tag{11}$$

where $\pi_i > 0$ for each $i = 1, ..., g$, $\sum_{i=1}^{g} \pi_i = 1$, and $f(x; \beta_i)$ is as in (5). Here, $\boldsymbol{\psi} = (\pi_1, ..., \pi_{g-1}, \beta_1, ..., \beta_g)^T$ is the parameter vector of (11). If $X$ has density function (11), then we say that $X$ arises from a $g$ component mixture of triangular distributions, or a $g$ component polygonal distribution in the language of Karlis and Xekalaki (2008).

Let $x_1, ..., x_n$ be a realization of the random sample $X_1, ..., X_n$ from a population characterized by the polygonal density function with parameter $\boldsymbol{\psi}_0$. It is clear that the log-likehood function of the sample,

$$\begin{aligned}
\ell_n(\boldsymbol{\psi}) &= \sum_{j=1}^{n} \log \sum_{i=1}^{g} \pi_i f(x; \beta_i) \\
&= \sum_{j=1}^{n} \log \sum_{i=1}^{g} \pi_i \min\{\beta_i x_j, 2\beta_i(1-x_j)/(\beta_i - 2)\}, \quad (12)
\end{aligned}$$

is difficult to maximize due to its log-of-sums form. Using the results from Section 3, we now propose an MM algorithm for the maximization of (12), in order to obtain an ML estimate $\hat{\boldsymbol{\psi}}_n$.

## 4.1 MM Algorithm

In order to construct an MM algorithm for the maximization of (12), we require the following result from Zhou and Lange (2010).

**Fact 4.** *If $\Theta = [0, \infty)^d$, then $\mathcal{L}(\boldsymbol{\theta}) = \log\left(\sum_{i=1}^d \theta_i\right)$ can be minorized by*

$$\mathcal{U}(\boldsymbol{\theta}; \boldsymbol{\psi}) = \sum_{i=1}^d \tau_i(\boldsymbol{\psi}) \log(\theta_i) - \sum_{i=1}^d \tau_i(\boldsymbol{\psi}) \log \tau_i(\boldsymbol{\psi}),$$

*where $\tau_i(\boldsymbol{\psi}) = \psi_i / \sum_{j=1}^d \psi_j$.*

Conditioned on the $r$th iterate $\boldsymbol{\psi}^{(r)}$ and using Fact 4, we can minorize (12) by

$$\begin{aligned}
\mathcal{U}_1\left(\boldsymbol{\psi}; \boldsymbol{\psi}^{(r)}\right) &= \sum_{i=1}^g \sum_{j=1}^n \tau_{ij}^{(r)} \log \pi_i - \sum_{i=1}^g \sum_{j=1}^n \tau_{ij}^{(r)} \log \tau_{ij}^{(r)} \\
&+ \sum_{i=1}^g \sum_{j=1}^n \tau_{ij}^{(r)} \min\{\beta_i x_j, 2\beta_i(1-x_j)/(\beta_i - 2)\},
\end{aligned} \quad (13)$$

where $\tau_{ij}^{(r)} = \pi_i^{(r)} f\left(x_j; \beta_i^{(r)}\right) / f\left(x_j; \boldsymbol{\psi}^{(r)}\right)$, by making the substitution $\theta_i = \pi_i f(x_j; \beta_i)$ for each $i$ and $j$. Since log is a concave function, a direct application of Proposition 1 yields the following result.

**Proposition 2.** *Conditioned on the $r$th iterate of the MM algorithm, both the minorizer (13) and log-likelihood function (12) can be minorized by*

$$\begin{aligned}
\mathcal{U}\left(\boldsymbol{\psi}; \boldsymbol{\psi}^{(r)}\right) &= \sum_{i=1}^g \sum_{j=1}^n \tau_{ij}^{(r)} \log \pi_i - \sum_{i=1}^g \sum_{j=1}^n \tau_{ij}^{(r)} \log \tau_{ij}^{(r)} \\
&\sum_{i=1}^g \sum_{j=1}^n \tau_{ij}^{(r)} \log u_2^{(r)}(x_j; \beta_i) - 2 \log n,
\end{aligned} \quad (14)$$

*and $\mathcal{U}\left(\boldsymbol{\psi}; \boldsymbol{\psi}^{(r)}\right)$ is a concave function in $\boldsymbol{\psi}$.*

It is well known that the unique global maximizer of $\sum_{i=1}^g \sum_{j=1}^n \tau_{ij}^{(r)} \log \pi_i$ over the unit simplex is $\pi_i^* = n^{-1} \sum_{j=1}^n \tau_{ij}^{(r)}$ (cf. McLachlan and Peel (2000, Ch. 2.8.2)). Furthermore, $\mathcal{U}\left(\boldsymbol{\psi}; \boldsymbol{\psi}^{(r)}\right)$ is concave and linearly separable in its parameter elements.

The MM algorithm can be summarized as follows. First, initialize the algorithm by some $\boldsymbol{\psi}^{(0)}$. At the $(r+1)$th iteration of the algorithm, set $\pi_i^{(r+1)} = n^{-1}\sum_{j=1}^n \tau_{ij}^{(r)}$, and $\beta_i^{(r+1)} = \beta^*$ where $\beta^*$ maximizes $\sum_{j=1}^n \tau_{ij}^{(r)} \log u_2^{(r)}(x_j; \beta)$, for each $i = 1, ..., g$.

The algorithm is iterated until $\ell\left(\boldsymbol{\psi}^{(r+1)}\right) - \ell\left(\boldsymbol{\psi}^{(r)}\right) < \epsilon$, whereupon we terminate the algorithm and declare the final iterate the ML estimate $\hat{\boldsymbol{\psi}}_n$. Like in Section 3, if we define $\boldsymbol{\psi}^{(\infty)} = \lim_{r\to\infty} \boldsymbol{\psi}^{(r)}$ to be a finite limit point of the MM algorithm, and note that the algorithm is an SLM algorithm, then we get the following result by application of Razaviyayn et al. (2013, Thm. 1).

**Theorem 2.** *If $\boldsymbol{\psi}^{(\infty)}$ is a limit point of the MM algorithm, for some initialization $\boldsymbol{\psi}^{(0)}$, then $\boldsymbol{\psi}^{(\infty)}$ is a local maximizer or inflection point of (12).*

Thus, like the algorithm from Section 3, the MM algorithm for the maximization of (12), again, monotonically increases the likelihood evaluations and is globally convergent. Unfortunately it is not simple to derive a method of moments estimator for the polygonal density function parameters, and thus a good initialization $\boldsymbol{\psi}^{(0)}$ is not easily attainable. Furthermore, it is known that the log-likelihood function of a mixture distribution is generally multimodal.

One technique for obtaining good initializations is that of McLachlan and Peel (2000, Sec. 2.12.2). Using the clustering interpretation of a mixture model, we can assign arbitrary labels to the data according to some predetermined prior probabilities $\pi_1, ..., \pi_{g-1}$ (e.g. if we let $\pi_1, ..., \pi_{g-1} = 1/g$, then for each $j$ we sample $i^* \in \{1, ..., g\}$ uniformly, and set $\tau_{ij}^{(0)} = 1$ if $i = i^*$ and 0 otherwise). Using the labeled data, we proceed to compute the estimate $\boldsymbol{\psi}^{(1)}$ and the respective log-likelihood $\ell_n\left(\boldsymbol{\psi}^{(1)}\right)$. We repeat this process multiple times and note the estimate and log-likelihood values from each repetition. The estimate that yields the highest log-likelihood value is considered the best initialization among the randomizations.

## 4.2 Time Complexity

The MM algorithm for the polygonal distribution with $g$ components is, in effect, maximizing $g$ triangular distribution weighted-log-likelihoods in each iteration. As such, using the same notation as in Section 3.4, the time complexity of the MM algorithm for maximizing the log-likelihood of a $g$ component polygonal distribution is of order $O\left(gnA_N^{[n]} A_{MM}^{[n]}\right)$.

In the EM algorithm of Karlis and Xekalaki (2008), it is suggested that the subproblems of maximizing $\sum_{j=1}^n \tau_{ij}^{(r)} f(x_j; \beta_i)$ be solved using exhaustive enumeration over all observations to obtain the updates $\beta_i^{(r+1)}$, for each $i$, at each iteration. The algorithm suggested in Karlis and Xekalaki (2008) therefore has a complexity of order $O\left(gn^2 A_{EM}^{[n]}\right)$, where $A_{EM}^{[n]}$ is the average number of EM iterations required to achieve a convergence level of $\epsilon$. Thus, the MM algorithm would be competitive to the current method if $A_N^{[n]} A_{MM}^{[n]}$ grows slower than $nA_{EM}^{[n]}$.

## 5 Numerical Simulations

We present two sets of numerical simulations, S1 and S2, in order to assess the performance of our MM algorithms. In S1, we compare the MM algorithm from Section 3 to the triangular distribution ML estimation method of Oliver (1972). In S2, we compare the MM algorithm from Section 3 to the polygonal distribution EM algorithm of Karlis and Xekalaki (2008).

All of our simulations are performed in the $R$ statistical programming environment (R Core Team, 2013), on an Intel Core i7-2600 CPU running at 3.40 GHz with 16 GB of internal RAM. In particular, we utilize the *sort* function in $R$, which is an implementation of the *quick sort* algorithm (see Cormen et al. (2002, Ch. 7) for details), for the sorting of samples necessary in the computation of (3). Furthermore, all computations of log-likelihood values, minorizers, and derivatives are performed using functions that are programmed in *Rcpp* and *Rcpparmadillo* (Eddelbuettel, 2013). The convergence thresholds for the EM algorithm, MM algorithm, and Newton algorithm are all set to $\epsilon = 10^{-3}$, and function timing is conducted using the *proc.time* function in $R$.

### 5.1 Simulation S1

In S1, we simulate $n \in \{5, 10, 20, 50, 100, 500, 1000, 10000, 100000, 1000000\}$ observations from triangular distributions with $\beta_0 \in \{4, 5, 20/3, 10, 20\}$, as per the simulations from Huang and Shen (2007). For each combination of $n$ and $\beta$, we compute (3) and note the time taken in seconds. Additionally, we apply the MM algorithm from Section 3 and note the time taken, as well as the number of MM iterations and Newton iterations that are required. Each combination is repeated 1000 times and the average times and number of iterations are reported in Table 1.

From Table 1, we see that the computation times of the MM algorithm are not competitive with the times taken to compute(3), in the range of scenarios assessed. At best, the MM algorithm is 1.84 times slower than (3) ($n = 10$ and $\beta_0 = 5$), and at worst, the MM algorithm is almost 300 times slower ($n = 1000$ and $\beta_0 = 4$).

We plot the product of $A_{MM}^{[n]}$ and $A_N^{[n]}$ against $\log_{10} n$, for each $\beta_0$, in Figure 2. Upon inspection, it is observable that the products are decreasing in $\log_{10} n$. Further, as $n$ increases, the product appears to reach a small asymptotic constant in each case. Thus, based on the conclusions from Section 3.4, the MM algorithm may become more competitive to (3) as $n$ increases.

In Figure 3, we plot the logarithm of the average time taken to compute the ML using the MM algorithm divided by the average time taken using (3) against $\log_{10} n$, for each $\beta_0$. It is observable that in each case, the ratio increases up until $n = 500$ or $n = 1000$, and then decreases as $n$ increases. This result provides further evidence that the MM algorithm may become more competitive to (3) for larger $n$.

Further results regarding the the accuracy of the ML estimates that are computed using either the MM algorithm or (3) are provided in Appendix A.

Table 1: Simulation results for S1. The rows Oliver and MM report the average time taken to compute the ML estimate using the respective methods, in seconds. The rows $A_{MM}^{[n]}$ and $A_N^{[n]}$ report the average number of MM and Newton iterations taken. All averages are taken over 1000 repetitions.

| $\beta_0$ | $n$ | 5 | 10 | 20 | 50 | 100 | 500 | 1000 | 10000 | 100000 | 1000000 |
|---|---|---|---|---|---|---|---|---|---|---|---|
| 4 | Oliver | 0.00007 | 0.00007 | 0.00005 | 0.00019 | 0.00011 | 0.00013 | 0.00014 | 0.00203 | 0.02125 | 0.26572 |
|   | MM | 0.00023 | 0.00036 | 0.00038 | 0.00101 | 0.00158 | 0.00578 | 0.00858 | 0.03511 | 0.38384 | 3.97548 |
|   | $A_{MM}^{[n]}$ | 3.553 | 3.194 | 2.596 | 2.235 | 2.129 | 2.287 | 2.646 | 7.733 | 12.627 | 13.053 |
|   | $A_N^{[n]}$ | 40.724 | 23.826 | 23.869 | 25.907 | 24.593 | 18.872 | 11.365 | 1.554 | 1.000 | 1.000 |
| 5 | Oliver | 0.00010 | 0.00013 | 0.00005 | 0.00005 | 0.00013 | 0.00011 | 0.00015 | 0.00186 | 0.02167 | 0.26583 |
|   | MM | 0.00029 | 0.00024 | 0.00033 | 0.00092 | 0.0018 | 0.00805 | 0.01163 | 0.0434 | 0.39342 | 4.36454 |
|   | $A_{MM}^{[n]}$ | 3.434 | 2.989 | 2.515 | 2.337 | 2.180 | 1.945 | 2.143 | 5.480 | 13.006 | 14.596 |
|   | $A_N^{[n]}$ | 50.328 | 32.188 | 25.007 | 28.320 | 32.209 | 30.392 | 20.687 | 2.892 | 1.010 | 1.000 |
| 20/3 | Oliver | 0.00019 | 0.00009 | 0.00011 | 0.00014 | 0.00008 | 0.00012 | 0.00035 | 0.0023 | 0.02169 | 0.26682 |
|   | MM | 0.00039 | 0.00043 | 0.00066 | 0.00145 | 0.00246 | 0.01011 | 0.01937 | 0.0731 | 0.41568 | 4.93723 |
|   | $A_{MM}^{[n]}$ | 3.384 | 3.013 | 2.652 | 2.441 | 2.193 | 1.983 | 1.957 | 3.981 | 12.830 | 16.517 |
|   | $A_N^{[n]}$ | 75.872 | 49.292 | 37.250 | 37.605 | 39.868 | 40.978 | 37.492 | 6.954 | 1.095 | 1.000 |
| 10 | Oliver | 0.00007 | 0.00014 | 0.00009 | 0.00013 | 0.00013 | 0.00025 | 0.00037 | 0.00226 | 0.02211 | 0.26784 |
|   | MM | 0.00038 | 0.00072 | 0.00109 | 0.00234 | 0.00352 | 0.01658 | 0.03213 | 0.15409 | 0.57734 | 6.10410 |
|   | $A_{MM}^{[n]}$ | 3.173 | 2.965 | 2.774 | 2.557 | 2.372 | 2.111 | 2.090 | 2.538 | 10.822 | 20.426 |
|   | $A_N^{[n]}$ | 87.219 | 76.464 | 67.396 | 66.100 | 51.434 | 59.792 | 58.826 | 23.483 | 1.916 | 1.001 |
| 20 | Oliver | 0.00009 | 0.00018 | 0.00014 | 0.00008 | 0.00014 | 0.00014 | 0.00024 | 0.00187 | 0.02148 | 0.25625 |
|   | MM | 0.00064 | 0.00110 | 0.00182 | 0.00408 | 0.00717 | 0.03536 | 0.06947 | 0.44876 | 1.69513 | 8.73229 |
|   | $A_{MM}^{[n]}$ | 3.348 | 3.194 | 2.926 | 2.732 | 2.618 | 2.341 | 2.258 | 2.027 | 4.807 | 24.595 |
|   | $A_N^{[n]}$ | 133.337 | 114.601 | 102.336 | 108.354 | 106.633 | 116.912 | 117.251 | 85.926 | 13.589 | 1.219 |

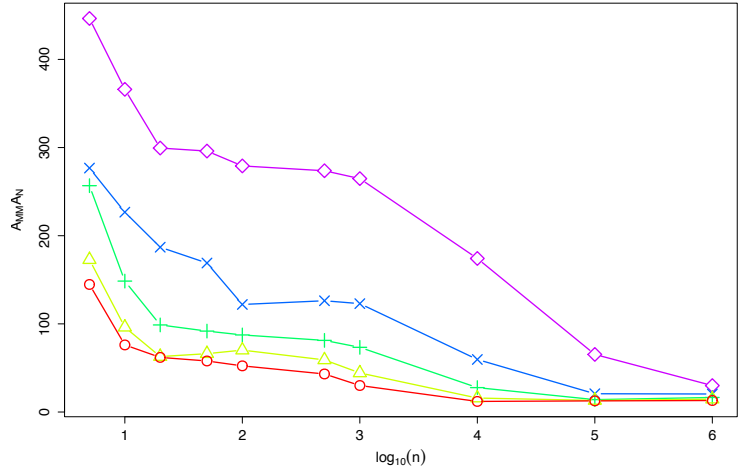

Figure 2: Product of $A_{MM}^{[n]}$ and $A_N^{[n]}$ versus $\log_{10} n$, in S1. Circles, triangles, pluses, crosses, and diamonds indicate the scenarios $\beta_0 = 4, 5, 20/3, 10, 20$, respectively.

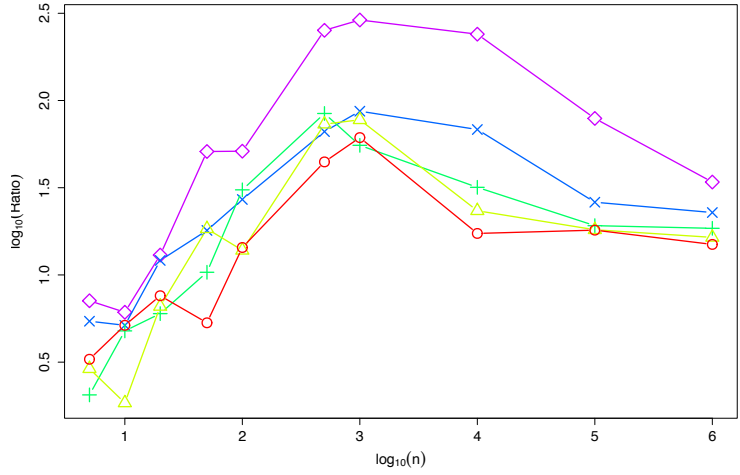

Figure 3: Log-ratio of the average computation time for the MM algorithm and the average computation time for (3) versus $\log_{10} n$, in S1. Circles, triangles, pluses, crosses, and diamonds indicate the scenarios $\beta_0 = 4, 5, 20/3, 10, 20$, respectively.

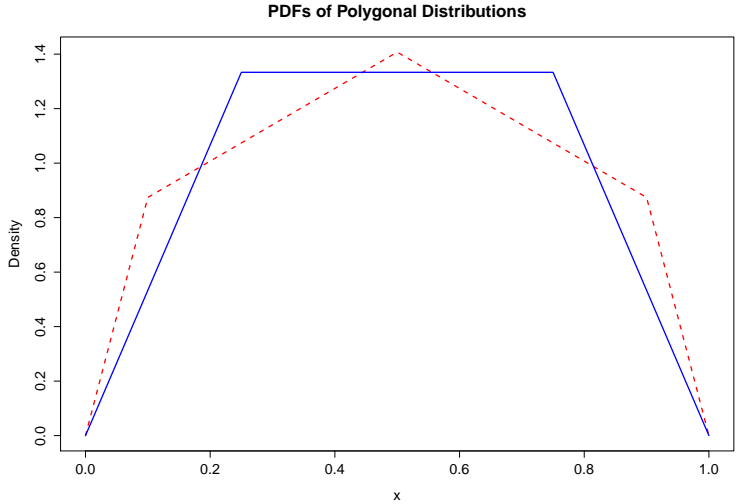

Figure 4: Plots of the Case 1 (solid) and Case 2 (dashed) polygonal probability density functions.

Results regarding the sensitivity of our conclusions to changes in $\epsilon$ can be found in Appendix B.

## 5.2 Simulation S2

In S2, we simulate $n \in \{5, 10, 20, 50, 100, 500, 1000\}$ observations from two polygonal distributions, Cases 1 and 2, that were suggested by Karlis and Xekalaki (2008). In Case 1, $g = 2$ and

$$\boldsymbol{\psi}_0 = (\pi_{01}, \beta_{01}, \beta_{02})^T = \left(\frac{1}{2}, \frac{8}{3}, 8\right)^T,$$

and in Case 2, $g = 3$ and

$$\boldsymbol{\psi}_0 = (\pi_{01}, \pi_{02}, \beta_{01}, \beta_{02}, \beta_{03})^T = \left(\frac{1}{3}, \frac{1}{3}, \frac{20}{9}, 4, 20\right)^T.$$

Probability density functions for the two cases are displayed in Figure 2. For each case and each $n$, we compute the ML estimate via the EM algorithm of Karlis and Xekalaki (2008) and note the time taken in seconds, as well as the number of EM iterations taken. Additionally, we apply the MM algorithm from Section 4 and note the time taken, as well as the number of MM iterations that are required. Each combination is repeated 1000 times and the average times and number of iterations are reported in Table 2.

From Table 2, we see that the MM algorithm is faster than the EM algorithm in all cases except for in Case 1 with $n = 5$, where the MM algorithm is 1.03 times slower than the EM algorithm. In Figure 5, we plot the logarithm of the average computation-time ratio for the MM algorithm versus the EM algorithm, for the two simulation cases. Upon inspection of Figure 5, we see that the ratio is decreasing as $n$ increases, for both cases, indicating that the MM algorithm scales better than the EM algorithm, with respect to $n$.

Table 2: Simulation results for S2. The rows EM and MM report the average time taken to compute the ML estimate using the respective methods, in seconds. The rows $A_{EM}^{[n]}$ and $A_{MM}^{[n]}$ report the average number of EM and MM iterations taken. All averages are taken over 1000 repetitions.

| Case | $n$ | 5 | 10 | 20 | 50 | 100 | 500 | 1000 |
|---|---|---|---|---|---|---|---|---|
| 1 | EM | 0.00097 | 0.00285 | 0.00562 | 0.01809 | 0.03889 | 0.29362 | 0.85880 |
|   | MM | 0.00100 | 0.00172 | 0.00207 | 0.00537 | 0.00912 | 0.04291 | 0.08671 |
|   | $A_{EM}^{[n]}$ | 6.219 | 9.494 | 11.673 | 15.164 | 16.244 | 16.944 | 17.513 |
|   | $A_{MM}^{[n]}$ | 9.248 | 13.646 | 13.182 | 15.921 | 17.329 | 23.318 | 30.001 |
| 2 | EM | 0.00206 | 0.00596 | 0.01262 | 0.04252 | 0.11752 | 1.43063 | 5.12175 |
|   | MM | 0.00154 | 0.00196 | 0.00472 | 0.01053 | 0.02176 | 0.18104 | 0.49679 |
|   | $A_{EM}^{[n]}$ | 9.505 | 13.285 | 17.311 | 24.658 | 33.503 | 56.824 | 72.536 |
|   | $A_{MM}^{[n]}$ | 11.436 | 17.772 | 26.304 | 39.620 | 49.130 | 77.511 | 98.241 |

The results from Section 5.1 indicate that the average number of Newton steps $A_N^{[n]}$ appears to decrease as a function of $n$. Thus, from the analysis of Section 4.2, the MM algorithm would be an improvement over the EM algorithm if $A_{MM}^{[n]}$ and $A_{EM}^{[n]}$ both grow at similar rates, with respect to $n$. In Figure 6, we plot $A_{MM}^{[n]}$ and $A_{EM}^{[n]}$ as a function of $\log_{10} n$, for the two simulation cases. Figure 6 indicates that the rate of growth of $A_{MM}^{[n]}$ and $A_{EM}^{[n]}$ appear to be linear in $\log_{10} n$, for both simulation cases. Furthermore, the rates of growth of $A_{MM}^{[n]}$ and $A_{EM}^{[n]}$ appear to be similar in both cases, which indicates that the MM algorithm improves upon the EM algorithm in the scenarios studied. Further results regarding the the accuracy of the ML estimates that are computed using either the MM or EM algorithms are provided in Appendix C.

## 6 Conclusions

In this article, we considered the ML estimation of the triangular distribution via an MM algorithm. The MM algorithm was constructed via a novel parametrization of the triangular distribution. The constructed algorithm was shown to have desirable numerical properties, such as monotonicity in likelihood evaluations, and convergence to a saddle point or local maximum of the log-likelihood function.

Using the novel triangular distribution parametrization, we also proposed an MM algorithm for the ML estimation of the polygonal distribution. This algorithm retains the desirable numerical properties that were shown for the triangular distribution MM algorithm.

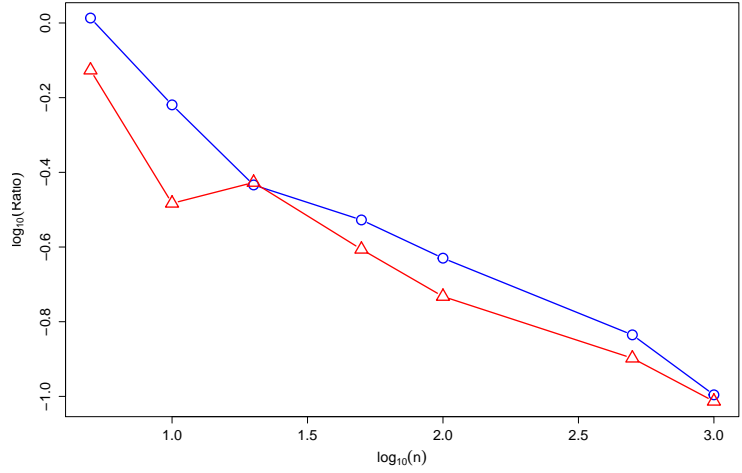

Figure 5: Log-ratio of the average computation time for the MM and EM algorithms versus $\log_{10} n$, in S2. Circles and triangles indicate Cases 1 and 2, respectively.

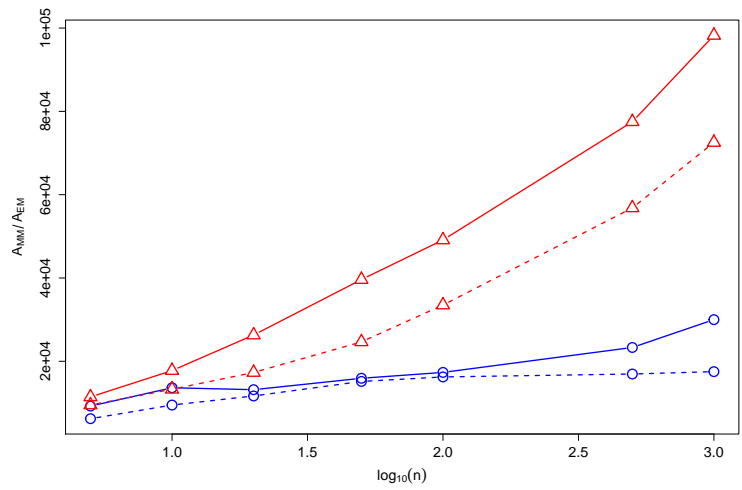

Figure 6: Average number of EM and MM iterations versus the $\log_{10} n$, from S2. Solid and dashed lines refer to the MM and EM algorithm results, respectively. Circles and triangles refer to Cases 1 and 2 results, respectively.

To assess our algorithms, we performed a set of numerical simulations. The MM algorithm for the triangular distribution appears to be not competitive with the established method, in the scenarios that were assessed. However, analysis of required numbers of iterations indicate that the MM algorithm scales well with increases in the number of observations. The MM algorithm for the polygonal distribution is shown to be faster than the currently used EM algorithm, in the scenarios that were assessed.

We believe that this article introduces an interesting approach to the ML estimation of piecewise linear density functions. In the future, it would not be difficult to extend our current methodology to the ML estimation of other piecewise densities, such as the trapezoidal densities of van Dorp and Kotz (2003), or other distributions that are defined by piecewise density functions on an interval support such as the generalized trapezoidal and two-sided distributions that are proposed in Kotz and van Dorp (2004). Such distributions have applications in modeling axial burnup distributions in nuclear engineering and USA certificate deposit rate distributions in financial engineering (cf. Sections 5 and 6 of Kotz and van Dorp (2004), respectively).

Furthermore, the new estimation techniques that are discussed in this article may be applied to the estimation of piecewise density functions that are defined on compact two-dimensional or multidimensional supports. Although we know of no such densities currently, it is possible to construct pyramidal-type density functions using the concept of nodal basis functions in finite element analysis; see Sangalli et al. (2013) and Nguyen et al. (2016), for example.

# Appendices

## A. Accuracy of the ML Estimates in S1

For each $\beta_0$ and $n$, we simulate data from a triangular distribution as per Section 5.1. Using these data, we compute the squared error of the ML estimate computed using either the MM algorithm or (3) to the true parameter $\beta_0$. This process is repeated 1000 times and the MSE (mean-squared error) is calculated as the average squared error over the replicates, for each combination of $\beta_0$ and $n$. The results of these simulations are presented in Table 3.

Upon inspection of Table 3, we see that the MSE of the ML estimates are decreasing in $n$ as would generally be expected. This appears to be the case for both the estimates obtained via the MM algorithm and via (3), and across all values of $\beta_0$. The Ratio row of Table 3 appears to indicate that neither the MM algorithm nor (3) yield consistently more accurate estimates than the other.

## B. Sensitivity of Results to $\epsilon$ in S1

We now analyze whether the conclusions from Section 5.1 and Appendix A are sensitive to different values of $\epsilon$. To conduct the analysis, we repeat the simulation study from Section 5.1 and Appendix A with $\epsilon$ set to $10^{-2}$ and $10^{-4}$, instead. For brevity, we do produce plots of the product of $A_{MM}^{[n]}$ and $A_N^{[n]}$ against $\log_{10} n$ and plots of the logarithm of the average times taken to compute the ML using the MM algorithm divided by the average time taken using (3) against $\log_{10} n$, for each $\beta_0$ (as in Figure 2 and 3), instead of the full result tables (as in Table 1). The plots for the $\epsilon = 10^{-2}$ and $\epsilon = 10^{-4}$ cases are presented in Figures 7 and 8, and Figures 9 and 10, respectively. Tables 4 presents the MSE results for the algorithm using $\epsilon = 10^{-2}$ and $\epsilon = 10^{-4}$, respectively.

Table 3: MSE results for S1. The rows Oliver and MM report the MSEs of the ML estimates using the respective techniques. The MSEs are computed as average squared errors over 1000 repetitions. The Ratio row reports the ratio of the MM and Oliver rows. The notation $a$E$b$ equates to $a \times 10^b$.

| $\beta_0$ | $n$ | 5 | 10 | 20 | 50 | 100 | 500 | 1000 | 10000 | 100000 | 1000000 |
|---|---|---|---|---|---|---|---|---|---|---|---|
| 4 | Oliver | 3.08E+03 | 8.53E+02 | 1.59E+02 | 8.15E-01 | 2.12E-01 | 3.77E-02 | 1.76E-02 | 1.69E-03 | 1.67E-04 | 1.55E-05 |
|   | MM | 4.83E+02 | 3.09E+01 | 6.39E+00 | 9.94E-01 | 2.74E-01 | 4.54E-02 | 1.97E-02 | 1.67E-03 | 1.58E-04 | 1.49E-05 |
|   | Ratio | 0.16 | 0.04 | 0.04 | 1.22 | 1.29 | 1.20 | 1.12 | 0.98 | 0.94 | 0.96 |
| 5 | Oliver | 5.83E+02 | 5.04E+02 | 2.23E+02 | 3.38E+00 | 7.36E-01 | 8.74E-02 | 4.23E-02 | 3.85E-03 | 3.67E-04 | 3.80E-05 |
|   | MM | 4.63E+02 | 6.14E+01 | 8.53E+00 | 2.16E+00 | 7.54E-01 | 1.13E-01 | 4.88E-02 | 3.96E-03 | 3.47E-04 | 3.54E-05 |
|   | Ratio | 0.80 | 0.12 | 0.04 | 0.64 | 1.03 | 1.30 | 1.15 | 1.03 | 0.94 | 0.93 |
| 20/3 | Oliver | 4.83E+03 | 1.18E+03 | 4.93E+02 | 1.12E+01 | 2.06E+00 | 2.41E-01 | 1.18E-01 | 1.09E-02 | 1.06E-03 | 1.12E-04 |
|   | MM | 2.13E+03 | 3.54E+03 | 3.52E+01 | 7.49E+00 | 3.06E+00 | 3.00E-01 | 1.52E-01 | 1.21E-02 | 1.01E-03 | 1.02E-04 |
|   | Ratio | 4.41 | 3.00 | 0.07 | 0.67 | 1.48 | 1.25 | 1.29 | 1.11 | 0.96 | 0.92 |
| 10 | Oliver | 1.96E+05 | 2.50E+03 | 1.34E+03 | 1.15E+03 | 9.48E+00 | 1.01E+00 | 4.50E-01 | 4.41E-02 | 3.85E-03 | 4.17E-04 |
|   | MM | 1.82E+04 | 3.88E+03 | 2.18E+02 | 5.53E+01 | 8.21E+00 | 1.27E+00 | 5.40E-01 | 5.01E-02 | 3.43E-03 | 3.72E-04 |
|   | Ratio | 0.09 | 1.55 | 0.16 | 0.05 | 0.87 | 1.25 | 1.20 | 1.14 | 0.89 | 0.89 |
| 20 | Oliver | 8.57E+03 | 3.91E+03 | 1.52E+03 | 8.43E+02 | 7.74E+02 | 1.05E+01 | 5.10E+00 | 3.73E-01 | 3.72E-02 | 3.62E-03 |
|   | MM | 5.01E+03 | 2.02E+03 | 9.62E+02 | 4.43E+02 | 1.23E+02 | 1.31E+01 | 5.60E+00 | 4.58E-01 | 4.14E-02 | 2.83E-03 |
|   | Ratio | 0.59 | 0.52 | 0.63 | 0.53 | 0.16 | 1.24 | 1.10 | 1.23 | 1.11 | 0.78 |

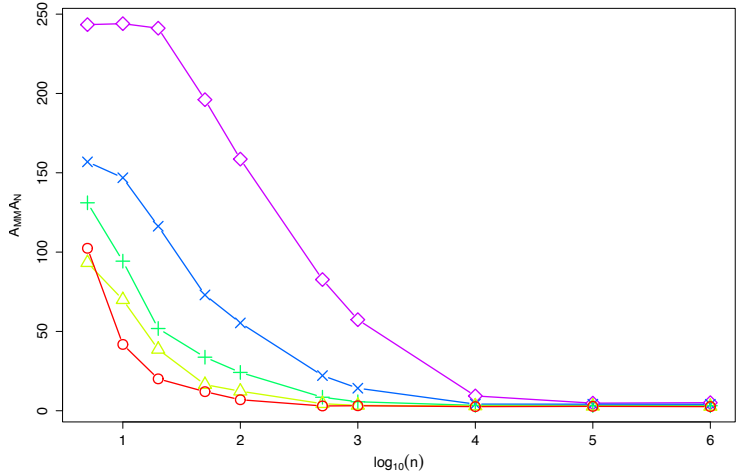

Figure 7: Product of $A_{MM}^{[n]}$ and $A_N^{[n]}$ versus $\log_{10} n$, in S1 with $\epsilon = 10^{-2}$. Circles, triangles, pluses, crosses, and diamonds indicate the scenarios $\beta_0 = 4, 5, 20/3, 10, 20$, respectively.

Upon inspection of Figures 7 and 9, we observe that the product of $A_{MM}^{[n]}$ and $A_N^{[n]}$ is decreasing for each value of $\beta_0$ and for both $\epsilon$ values. Figures 8 and 10 indicate that the average ratio of computation times is also a decreasing function of $n$, after $n = 1000$. Both of these results conform with the conclusions from Section 5.1. The results of Table 4 closely corresponds with those obtained in Appendix A. We note that the MSE is decreasing in $n$, and that there are no discernible structural relationships between the MSEs of the MM algorithm and those of (3).

## C. Accuracy of ML Estimates in S2

Using the two cases and values of $n$, from Section 5.2, we simulate data from polygonal distributions. Using these data, we compute the squared error of the ML estimate computed using either the EM algorithm or MM algorithm to the true parameter vector $\boldsymbol{\psi}_0$. This process is repeated 1000 times and the MSE (mean-squared error) is calculated as the average squared error over the replicates, for each element of the parameter vector in each of the case and each $n$. The results of these simulations are presented in Table 5.

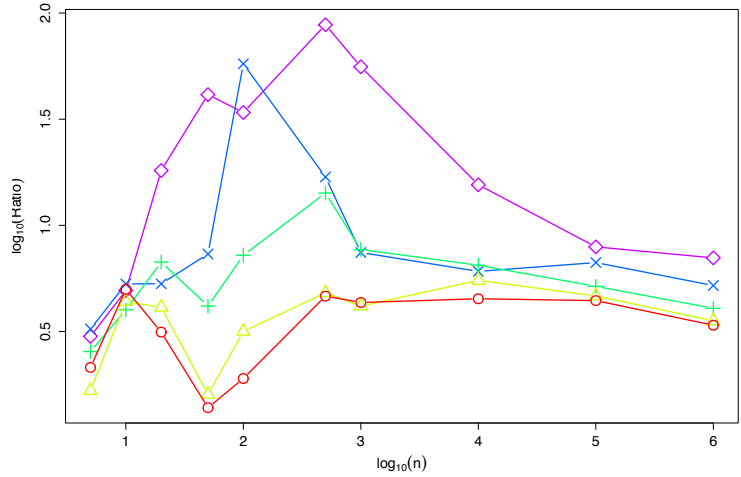

Figure 8: Log-ratio of the average computation time for the MM algorithm and the average computation time for (3) versus $\log_{10} n$, in S1 with $\epsilon = 10^{-2}$. Circles, triangles, pluses, crosses, and diamonds indicate the scenarios $\beta_0 = 4, 5, 20/3, 10, 20$, respectively.

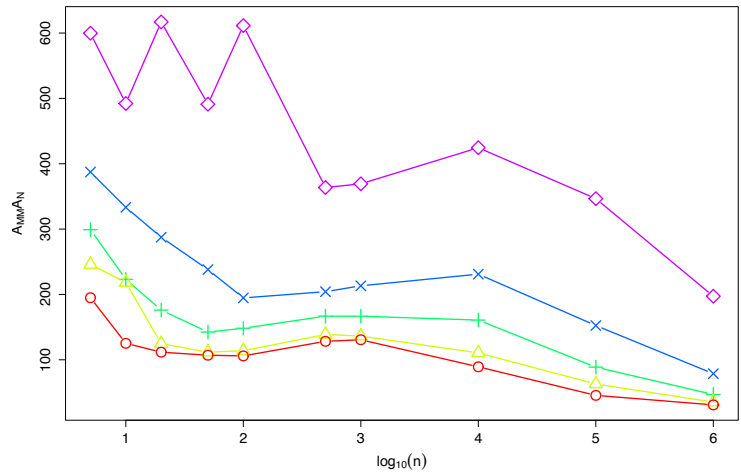

Figure 9: Product of $A_{MM}^{[n]}$ and $A_N^{[n]}$ versus $\log_{10} n$, in S1 with $\epsilon = 10^{-4}$. Circles, triangles, pluses, crosses, and diamonds indicate the scenarios $\beta_0 = 4, 5, 20/3, 10, 20$, respectively.

Table 4: Additional MSE results for S1. The rows $\epsilon = 10^{-2}$ and $\epsilon = 10^{-4}$ report the MSEs of the ML estimates using MM algorithm with the respective convergence criteria. The MSEs are computed as average squared errors over 1000 repetitions. The notation $aEb$ equates to $a \times 10^b$.

| $\beta_0$ | $n$ | 5 | 10 | 20 | 50 | 100 | 500 | 1000 | 10000 | 100000 | 1000000 |
|---|---|---|---|---|---|---|---|---|---|---|---|
| 4 | $\epsilon = 10^{-2}$ | 3.04E+03 | 2.10E+01 | 3.49E+00 | 1.05E+00 | 2.58E-01 | 3.36E-02 | 1.65E-02 | 1.78E-03 | 1.60E-04 | 1.70E-05 |
|   | $\epsilon = 10^{-4}$ | 2.43E+03 | 2.94E+01 | 6.32E+00 | 7.44E-01 | 3.83E-01 | 5.20E-02 | 2.59E-02 | 2.09E-03 | 1.84E-04 | 1.50E-05 |
| 5 | $\epsilon = 10^{-2}$ | 4.27E+02 | 5.29E+01 | 1.52E+01 | 1.56E+00 | 7.04E-01 | 7.81E-02 | 3.96E-02 | 3.98E-03 | 3.80E-04 | 3.68E-05 |
|   | $\epsilon = 10^{-4}$ | 1.18E+03 | 6.52E+01 | 9.18E+00 | 2.01E+00 | 6.79E-01 | 1.24E-01 | 5.27E-02 | 5.00E-03 | 4.51E-04 | 3.62E-05 |
| 20/3 | $\epsilon = 10^{-2}$ | 2.01E+03 | 1.94E+02 | 2.81E+01 | 4.56E+00 | 1.94E+00 | 2.52E-01 | 1.13E-01 | 9.45E-03 | 9.73E-04 | 1.00E-04 |
|   | $\epsilon = 10^{-4}$ | 1.36E+04 | 5.28E+03 | 5.17E+01 | 7.69E+00 | 2.85E+00 | 3.20E-01 | 1.47E-01 | 1.40E-02 | 1.18E-03 | 1.03E-04 |
| 10 | $\epsilon = 10^{-2}$ | 4.27E+02 | 5.29E+01 | 1.52E+01 | 1.56E+00 | 7.04E-01 | 7.81E-02 | 3.96E-02 | 3.98E-03 | 3.80E-04 | 3.68E-05 |
|   | $\epsilon = 10^{-4}$ | 1.70E+04 | 9.81E+03 | 1.73E+02 | 6.97E+01 | 8.02E+00 | 1.15E+00 | 5.71E-01 | 5.32E-02 | 4.68E-03 | 4.29E-04 |
| 20 | $\epsilon = 10^{-2}$ | 1.01E+04 | 2.59E+03 | 5.58E+02 | 3.55E+02 | 1.35E+02 | 1.10E+01 | 4.06E+00 | 3.05E-01 | 2.65E-02 | 2.28E-03 |
|   | $\epsilon = 10^{-4}$ | 9.03E+03 | 8.91E+03 | 9.00E+02 | 3.03E+02 | 1.84E+02 | 1.44E+01 | 5.45E+00 | 4.90E-01 | 4.52E-02 | 4.43E-03 |

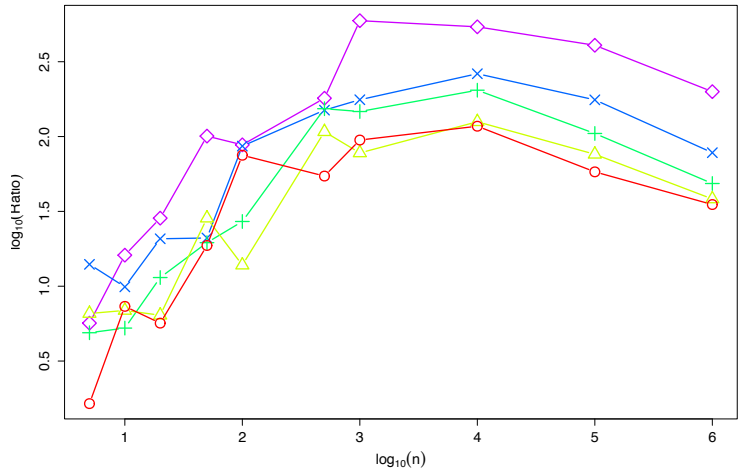

Figure 10: Log-ratio of the average computation time for the MM algorithm and the average computation time for (3) versus $\log_{10} n$, in S1 with $\epsilon = 10^{-4}$. Circles, triangles, pluses, crosses, and diamonds indicate the scenarios $\beta_0 = 4, 5, 20/3, 10, 20$, respectively.

Similarly to the results of Appendix A, we observe that the MSEs are decreasing in $n$, for both algorithms. However, there appears to be no evidence of either algorithm being superior in accuracy to the other.

# References

Boyd, S., Vandenberghe, L., 2004. Convex Optimization. Cambridge University Press, Cambridge.

Brent, R. P., 1971. An algorithm with guaranteed convergence for finding a zero of a function. Computer Journal 14, 422–425.

Cormen, T. H., Leiserson, C. E., Rivest, R. L., Stein, C., 2002. Introduction To Algorithms. MIT Press, Cambridge.

Eddelbuettel, D., 2013. Seamless R and C++ Integration with Rcpp. Springer, New York.

Forbes, C., Evans, M., Hastings, N., Peacock, B., 2011. Statistical Distributions. Wiley, New York.

Table 5: MSE results for S2. The rows EM and MM report the MSEs of the ML estimates using the respective techniques. For each parameter vector component, the MSEs are computed as average squared errors over 1000 repetitions. The Ratio row reports the ratio of the MM and EM rows. The notation $aEb$ equates to $a \times 10^b$.

| Case | Element | | n | 5 | 10 | 20 | 50 | 100 | 500 | 1000 |
|---|---|---|---|---|---|---|---|---|---|---|
| 1 | $\pi_1$ | EM | | 1.10E-01 | 8.81E-02 | 3.34E-02 | 8.07E-03 | 3.33E-03 | 2.95E-04 | 4.12E-06 |
| | | MM | | 4.06E-02 | 2.01E-02 | 9.03E-03 | 3.24E-03 | 1.23E-03 | 1.14E-03 | 1.10E-03 |
| | | Ratio | | 3.69E-01 | 2.28E-01 | 2.70E-01 | 4.01E-01 | 3.69E-01 | 3.86E+00 | 2.67E+02 |
| | $\beta_1$ | EM | | 7.43E+01 | 2.25E+01 | 1.76E+01 | 1.32E+01 | 3.51E+00 | 1.37E+00 | 2.00E-02 |
| | | MM | | 7.51E+01 | 7.15E+01 | 3.35E+00 | 1.17E+00 | 2.37E-01 | 1.72E-03 | 3.10E-04 |
| | | Ratio | | 1.01E+00 | 3.18E+00 | 1.90E-01 | 8.86E-02 | 6.75E-02 | 1.26E-03 | 1.55E-02 |
| | $\beta_2$ | EM | | 1.37E+00 | 6.87E-02 | 2.38E-02 | 8.98E-03 | 1.61E-03 | 6.06E-04 | 4.58E-06 |
| | | MM | | 1.78E-01 | 9.06E-02 | 8.26E-02 | 2.73E-02 | 1.18E-02 | 1.29E-04 | 1.79E-06 |
| | | Ratio | | 1.30E-01 | 1.32E+00 | 3.47E+00 | 3.04E+00 | 7.33E+00 | 2.13E-01 | 3.91E-01 |
| 2 | $\pi_1$ | EM | | 3.27E-03 | 2.05E-03 | 2.38E-04 | 1.73E-04 | 1.55E-04 | 9.24E-05 | 2.95E-05 |
| | | MM | | 5.69E-02 | 4.92E-02 | 4.19E-02 | 2.14E-02 | 9.48E-03 | 2.75E-03 | 2.49E-03 |
| | | Ratio | | 1.74E+01 | 2.40E+01 | 1.76E+02 | 1.24E+02 | 6.12E+01 | 2.98E+01 | 8.44E+01 |
| | $\pi_2$ | EM | | 2.27E-02 | 1.49E-02 | 5.42E-03 | 4.33E-03 | 3.97E-03 | 7.52E-04 | 4.37E-08 |
| | | MM | | 9.89E-02 | 8.08E-02 | 5.21E-02 | 3.54E-02 | 2.48E-02 | 2.26E-02 | 1.60E-02 |
| | | Ratio | | 4.36E+00 | 5.42E+00 | 9.61E+00 | 8.18E+00 | 6.25E+00 | 3.01E+01 | 3.66E+05 |
| | $\beta_1$ | EM | | 2.17E+02 | 1.88E+02 | 1.65E+02 | 1.32E+02 | 1.26E+02 | 1.17E+02 | 6.76E+01 |
| | | MM | | 1.16E+03 | 2.48E+02 | 2.19E+02 | 1.94E+02 | 1.20E+02 | 5.17E+01 | 5.14E+00 |
| | | Ratio | | 5.35E+00 | 1.32E+00 | 1.33E+00 | 1.47E+00 | 9.52E-01 | 4.42E-01 | 7.60E-02 |
| | $\beta_2$ | EM | | 2.02E+01 | 6.11E+00 | 1.64E+00 | 1.36E+00 | 4.98E-01 | 1.45E-02 | 9.06E-03 |
| | | MM | | 2.53E+00 | 2.25E-01 | 1.86E-01 | 1.38E-01 | 4.98E-02 | 3.89E-02 | 3.14E-02 |
| | | Ratio | | 1.25E-01 | 3.68E-02 | 1.13E-01 | 1.01E-01 | 1.00E-01 | 2.68E+00 | 3.47E+00 |
| | $\beta_3$ | EM | | 3.93E+01 | 1.81E+01 | 9.35E+00 | 8.67E+00 | 6.17E+00 | 2.83E+00 | 2.75E+00 |
| | | MM | | 1.14E+01 | 4.88E+00 | 4.62E+00 | 3.82E+00 | 2.50E+00 | 2.42E+00 | 1.70E+00 |
| | | Ratio | | 2.90E-01 | 2.70E-01 | 4.94E-01 | 4.41E-01 | 4.05E-01 | 8.55E-01 | 6.18E-01 |

Huang, J. S., Shen, P. S., 2007. More maximum likelihood oddities. Journal of Statistical Planning and Inference 137, 2151–2155.

Hunter, D. R., 2004. MM algorithms for generalized Bradley-Terry models. Annals of Statistics 32, 384–406.

Hunter, D. R., Li, R., 2005. Variable selection using MM algorithms. Annals of Statistics 33.

Johnson, N. L., Kotz, S., 1999. Non-smooth sailing or triangular distributions revisited after some 50 years. Statistician 48, 179–187.

Johnson, N. L., Kotz, S., Balakrishnan, N., 1995. Continuous Univariate Distributions. Vol. 2. Wiley, New York.

Karlis, D., Xekalaki, E., 2008. The polygonal distribution. In: Advances in Mathematical and Statistical Modeling. Birkhauser, Boston.

Kotz, S., van Dorp, J. R., 2004. Beyond Beta: Other Continuous Families of Distributions with Bounded Support and Applications. World Scientific, Singapore.

Lange, K., 2013. Optimization. Springer, New York.

McLachlan, G. J., Peel, D., 2000. Finite Mixture Models. Wiley, New York.

Nguyen, H. D., McLachlan, G. J., 2015. Maximum likelihood estimation of Gaussian mixture models without matrix operations. Advances in Data Analysis and Classification In Press.

Nguyen, H. D., McLachlan, G. J., 2016. Laplace mixture of linear experts. Computational Statistics and Data Analysis 93, 177–191.

Nguyen, H. D., McLachlan, G. J., Wood, I. A., 2016. Mixture of spatial spline regressions for clustering and classification. Computational Statistics and Data Analysis 93, 76–85.

Oliver, E. H., 1972. A maximum likelihood oddity. American Statistician 26, 43–44.

R Core Team, 2013. R: a language and environment for statistical computing. R Foundation for Statistical Computing.

Razaviyayn, M., Hong, M., Luo, Z.-Q., 2013. A unified convergence analysis of block successive minimization methods for nonsmooth optimization. SIAM Journal of Optimization 23, 1126–1153.

Sangalli, L. M., Ramsay, J. O., Ramsay, T. O., 2013. Spatial spline regression models. Journal of the Royal Statistical Society Series B 75, 681–703.

van Dorp, J. R., Kotz, S., 2003. Generalized trapezoidal distributions. Metrika 58, 85–97.

Wu, T. T., Lange, K., 2010. The MM alternative to EM. Statistical Science 25, 492–505.

Zhou, H., Lange, K., 2010. MM algorithms for some discrete multivariate distributions. Journal of Computational and Graphical Statistics 19, 645–665.

Zhou, H., Zhang, Y., 2012. EM vs MM: a case study. Computational Statistics and Data Analysis 56, 3909–3920.